\documentclass[aps,prl,twocolumn,floats,showpacs]{revtex4}
\usepackage{bm}
\usepackage{amsfonts}
\usepackage{graphicx,color}
\usepackage{epsfig}
\usepackage{amsmath}
\usepackage{setspace}

\newcommand{\beq}{\begin{equation}}
\newcommand{\eeq}{\end{equation}}
\newcommand{\bea}{\begin{eqnarray}}
\newcommand{\eea}{\end{eqnarray}}

\begin{document}
\title{Cold Atomic Collisions:
Coherent Control of Penning and Associative Ionization}
\author{Carlos A. Arango$^1$,
Moshe Shapiro$^{2,3}$, and Paul Brumer$^1$\\
$^1${\it Chemical Physics Theory Group,
Dept. of Chemistry, and Center for Quantum Information and Quantum
Control, University of Toronto, Toronto M5S3H6, Canada}\\
$^2${\it Dept. of Chemical Physics, The Weizmann Institute,
Rehovot 76100, Israel\\
$^3${\it Dept. of Chemistry, The University of British Columbia,
Vancouver V6T1Z1, Canada}}}

\date{\today}

\begin{abstract}

Coherent Control techniques are computationally applied to
cold ($1\,\mathrm{mK}<T<1\,\mathrm{K}$)
and ultracold  ($T<1\,\mu\mathrm{K}$) Ne$^*$($3s$,$^3P_2$) + Ar($^1S_0$)
collisions. We show that by using various
initial superpositions of the Ne$^*$($3s$,$^3P_2$)
$M=\{-2,-1,0,1,2\}$ Zeeman sub-levels it is possible to reduce
the Penning Ionization (PI) and Associative Ionization (AI)
cross sections by as much as four orders of magnitude. It is also possible to
drastically change the ratio of these two processes.
The results are based on combining, within the
``Rotating Atom Approximation'', empirical and {\it ab-initio}
ionization-widths.

\end{abstract}

\maketitle

Cold and ultracold atomic processes present a new
laboratory environment in which to explore and utilize the quantum nature
of matter. In this letter we show that the significance of quantum effects
in such systems permit unprecedented control over collisional processes.
In particular, we consider the theory and computational implementation of
the coherent control\cite{Shapiro03,Shapiro96,Brumer99,Rice} of absolute
and relative cross sections in the collision of metastable atoms $A^*$ and
ground state target atoms $B$. Such collisions can result in two main
channels: (1) the ionization of the target atom and the de-excitation of
the metastable species, i.e., Penning Ionization (PI) \cite{Siska93}, or
(2) Associative Ionization (AI), wherein the colliding partners form an
ionic dimer while emitting an energetic electron. Schematically,
\begin{equation}
({\rm AI})~~AB^+ + e^-  \leftarrow A^*+B \rightarrow A + B^+ + e^-,
~~({\rm PI})
\label{eqn.0.2}
\end{equation}
As an example, we consider the coherent control
of PI and AI resulting from collisions between
Ne$^*$($3s$,$^3P_2$) and Ar($^1S_0$) in the cold and ultracold regimes. 
Amongst other observations, the suppression of
these processes in favor of elastic scattering may well prove useful
for the production
of Bose-Einstein Condensates of excited states atoms.

A wealth of
experimental information and theoretical calculations on the
uncontrolled Ne$^*$ + Ar collisions is available and
the possibility of control of this system at thermal
energies\cite{Arango06} now exists.  
As we report below, the control achievable
in the sub mK regime is far more impressive.

{\it The Initial Superposition State:} Coherent Control is achieved by preparing
the colliding pair in an initial superposition of internal states, such as,
\begin{equation}\label{superposition}
|\psi\rangle=
{\mathrm e}^{{\mathrm i}{\bf K}\cdot{\bf R}_{\mathrm{CM}}
+{\mathrm i}{\bf k}\cdot{\bf r}}
|\phi_{\mathrm{Ar}}\rangle
\sum_Ma_M|\phi^M_{\mathrm{Ne^*}}\rangle~,
\end{equation}
where $|\psi_{\mathrm{Ar}}\rangle$ is the initial state of the Ar atom and
$|\phi^M_{\mathrm{Ne}^*}\rangle$,
are Ne$^*$  Zeeman sublevels, with
$M=\{-2,-1,0,1,2\}$
being the projection of
the Ne$^*$  electronic angular momentum on the space-fixed quantization axis.
$a_M$ are preparation coefficients, to be optimized to yield a desired
objective,
${\bf R}_{\mathrm{CM}}$ is the CM coordinate,
${\bf R}_{\mathrm{CM}}\equiv\left(m_{\mathrm{Ne}}{\bf r}^{\mathrm{Ne}}
+m_{\mathrm{Ar}}{\bf r}^{\mathrm{Ar}}\right)/
(m_{\mathrm{Ne}}+m_{\mathrm{Ar}})~,$
and
${\bf r}$ is the internuclear separation vector,
${\bf r}\equiv{\bf r}^{\mathrm{Ne}}-{\bf r}^{\mathrm{Ar}}~.$
The (Body-Fixed) momenta are given as,
${\bf K}\equiv{\bf k}^{\mathrm{Ne}}+{\bf k}^{\mathrm{Ar}}~,~~
{\bf k}\equiv\left(m_{\mathrm{Ar}}{\bf k}^{\mathrm{Ne}}
-m_{\mathrm{Ne}}{\bf k}^{\mathrm{Ar}}\right)/
(m_{\mathrm{Ne}}+m_{\mathrm{Ar}})~.$
Here ${\bf r}^{\mathrm{Ar}}$ and
${\bf k}^{\mathrm{Ar}}$ (${\bf r}^{\mathrm{Ne}}$ and
${\bf k}^{\mathrm{Ne}}$) denote the position and momentum
of the Ar (Ne$^*$) atom in the laboratory frame. Note that the fact
that the initial superposition state is comprised of  degenerate 
$M$ states, and that the collision partners are atoms, ensures that
the conditions for coherent control\cite{Brumer99} are satisfied.

The rates of the PI and AI processes mainly depend on $\lambda$, the
body-fixed (BF) projection of the electronic
angular momentum on ${\bf r}$, the interatomic axis.
It is therefore necessary to express the
$|\phi^M_{\mathrm{Ne^*}}\rangle$ states in terms of the
$|\phi^\lambda_{\mathrm{Ne^*}}\rangle$ BF states.
We adopt the ``Rotating Atom Approximation''\cite{Mori64}
according to which
the axis of quantization of the electrons faithfully follows
the internuclear separation vector. This establishes a 1:1 correspondence between
the $M$ values and the $\lambda$ values as
the atoms approach one another.
Hence, the (even parity) linear combination in the BF frame is written as,
\begin{equation}\label{1.7in1.1}
|\psi\rangle=|\phi_{\mathrm{Ar}}\rangle{\mathrm e}^{{\mathrm i}
{\bf K}\cdot{\bf R}_{\mathrm{CM}}+{\mathrm i}{\bf k}\cdot{\bf r}}
\sum_{\Omega=0}^2|\phi^\Omega_{\mathrm{Ne^*}}\rangle a_\Omega,
\end{equation}
where $\Omega\equiv|\lambda|$, and
(due to the assumed even parity) $a_{\Omega}\equiv(a_M+a_{-M})$.


{\it Scattering Theory:}
The basic formulae for our purposes are found in Refs.
\cite{Hickman76, Morgner78, Bieniek78} giving
the scattering amplitudes for PI and AI
based on O'Malley's theory of dissociative attachment
\cite{O'Malley66,O'Malley67}.
Prior to the collision the internuclear momentum vector ${\bf k}$ has magnitude $k$
and direction $\hat{\bf k}$.
After the collision its magnitude is $k_f$ and its direction is
$\hat{\bf k}_f$. Asymptotically the Penning electron departs along the
${\hat{\bf k}_\varepsilon}$ direction with energy $\varepsilon$.
The energy of the emitted electron is related to the collisional energy $E$
and the energy of the nuclei after the collision $E'$ by the conservation of
energy $E+\varepsilon_0=\varepsilon+E',$
with $\varepsilon_0=E_*-IE$ being the difference between the excitation energy of
the metastable Ne$^*$ atom and the ionization energy of the target Ar atom.

The scattering amplitude, which is exact within the Born-Oppenheimer
approximation, is given by \cite{Miller72,Morgner78,Hickman76}
\begin{equation}
\label{1.8in1}
f(\hat{\bf k}_f, \varepsilon,\hat{\bf k}_\varepsilon;{\bf k})=
-\frac{2M_r\rho_\varepsilon^{1/2}}{(4\pi\hbar)^2}
\left(\frac{k_f}{k}\right)^{1/2}\left<\psi_\varepsilon
\right|V_{\varepsilon,\hat{\bf k}_\varepsilon}|\psi_d\rangle,
\end{equation}
where $M_r$ is the reduced mass of the nuclei and $\rho_\varepsilon$ is the
density of electronic continuous states. $\psi_d({\bf r})$ is the incoming
wave function calculated on the optical potential
$V_*(r)-\tfrac{\mathrm i}{2}\Gamma(r)$ and $\psi_\varepsilon({\bf r})$
describes the system on the exit
channel $V_+(r)$. The electronic part is completely
included in $V_{\varepsilon,\hat{\bf k}_\varepsilon}({\bf r})$, which is the
probability amplitude for the emission of an electron with $\varepsilon$,
and $\hat{\bf k}_\varepsilon$.

Partial wave expansions  of $\psi_\varepsilon({\bm r})$,
$V_{\varepsilon,\hat\varepsilon}({\bm r})$, and $\psi_d({\bm r})$, and the
evaluation of the integral over $\bm r$ will give, for the special case
when the space fixed $z$-axis is along $\bm k,$
$$
f=\frac{\pi^{1/2}}{{\mathrm i}k}\sum_{\ell,\mu,l,l'}
{\mathrm i}^{l-l'}(2l+1)(2l'+1)^{1/2}
\begin{pmatrix}
l' & \ell & l \\ 0 & 0 & 0
\end{pmatrix}
\begin{pmatrix} l' & \ell & l \\ -\mu & \mu & 0
\end{pmatrix}
$$
\begin{equation}\label{1.8in18}
S^l_{l'\ell}(\varepsilon)
Y_{l'-\mu}(\hat{\bf k}_f)Y_{\ell\mu}({\hat{\bf k}_\varepsilon}),
\end{equation}
with the partial-wave $S$-matrix in terms of the phase shifts $\delta^l$
and $\delta_f^{l'}$ of the radial partial wave components $\psi^l_d$ and
$\psi^{l'}$
\begin{equation}\label{1.8in16}
S^l_{l'\ell}(\varepsilon)=-{\mathrm i}\frac{4M_r\rho_\varepsilon^{1/2}}
{\hbar^2}{\mathrm e}^{{\mathrm i}(\delta^l+\delta_f^{l'})}
\langle\psi^{l'}_\varepsilon|V_{\varepsilon\ell}|\psi^l_d\rangle.
\end{equation}
where $V_{\varepsilon\ell}(r)\approx\alpha_\ell\left[\Gamma(r)/2\pi\right]^{1/2}.$

For crossed beams, in the rotating atom approximation, we find that the
scattering amplitude for a linear superposition $\bm a_\Omega$ can be written as
\begin{equation}\label{1.7in3.2}
f_a({\hat k}_f, \varepsilon,\hat{\bf k}_\varepsilon,k)=
\sum_\Omega{a_\Omega f_\Omega({\hat k}_f, \varepsilon,\hat{\bf k}_\varepsilon,k)},
\end{equation}
with $f_\Omega({\hat k}_f, \varepsilon,\hat{\varepsilon};k,\Omega)$
given by equation \eqref{1.8in18} and $S^{l}_{l'\ell}(\varepsilon)$ replaced by
\begin{equation}
\label{1.23in4}
S^{l,\Omega}_{l'\ell}(\varepsilon)=-2{\mathrm i}
\frac{2M_r\rho_\varepsilon^{1/2}}{\hbar^2}
{\mathrm e}^{{\mathrm i}(\delta^l+\delta_f^{l'})}
\langle\psi^{l'}_\varepsilon|V^\Omega_{\varepsilon\ell}|\psi^l_d\rangle,
\end{equation}
for PI, and
\begin{equation}
\label{1.23in4a}
S^{l,\Omega}_{l'\ell}(\varepsilon)=-2{\mathrm i}
\left(\frac{2M_r\pi\rho_\varepsilon}{\hbar^2}\right)^{\tfrac{1}{2}}
{\mathrm e}^{{\mathrm i}\delta^l}
\langle\psi_{v'l'}|V^\Omega_{\varepsilon\ell}|\psi^l_d\rangle,
\end{equation}
for AI\cite{Siska93}. The $V^\Omega_{\varepsilon\ell}(r)$ matrix elements
are related to the ionization widths, $\Gamma_\Omega(r)$ as
\cite{Hickman76,Morgner78,Siska93},
$V^\Omega_{\varepsilon\ell}(r)\approx\alpha_\ell[\Gamma_\Omega(r)/
{2\pi}]^{1/2}$. The $\Gamma_\Omega$ are obtained as in Ref.
\cite{Arango06}, using methods in
\cite{Gregor81,Verheijen86,Kerstel88,OpdeBeek97}.

The differential cross section for PI is obtained by squaring
the scattering amplitude Eq. \eqref{1.8in18}.
In the rotating atom approximation we obtain
\begin{equation}\label{1.7in3.9}
\sigma_q(\hat{\bf k}_f,\varepsilon,\hat{\bf k}_\varepsilon; k,{\bf a}_\Omega)=
\left|\sum_\Omega a_\Omega f_{q\Omega}(\hat{\bf k}_f, \varepsilon,
\hat{\bf k}_\varepsilon;k)\right|^2,
\end{equation}
where $q$ indicates the exit channel. There are three possible exit
channels characterized by the electronic state of the products:
$X\phantom{.}^2\Sigma^+_{1/2}$, $A_1\phantom{.}^2\Pi_{3/2}$,
and $A_2\phantom{.}^2\Sigma_{1/2}$. The entrance channel
optical potential are connected
with each of the exit channels using Morgner's $\Gamma$
splitting \cite{Morgner82, Morgner85}.
We use optical potentials for the entrance channel derived directly
from experiment\cite{Gregor81,Verheijen86}.  Since data is available only for
scattering experiments at thermal energies
(above 1\,K), the optical potentials
do not include effects associated with very slowly moving atoms
(e.g., hyperfine interactions).

The sum Eq. \eqref{1.7in3.9} can be expanded to give
\begin{equation}\label{1.7in3.10}
\sigma_q(\hat{\bf k}_f,\varepsilon,\hat{\bf k}_\varepsilon; k,{\bf a}_\Omega)=
\sum_{\Omega\Omega'}a^*_{\Omega'}a_{\Omega}\sigma_{q\Omega'\Omega}
\end{equation}
with $\sigma_{q\Omega'\Omega}=f^*_{q\Omega'}f_{q\Omega}$. Since we are
interested in the total ionization cross section we sum over all
the exit channels, and integrate over the solid angles
$\hat{\bf k}_\varepsilon$
and $\hat {\bf k}_f$, and over the emitted electron energy $\varepsilon$.
It is easy to see from the form of the scattering amplitude that the
integration over the solid angles yields a Kronecker delta function, thus
simplifying the expressions for the $f^*_{q\Omega'}f_{q\Omega}$ products to:
\begin{equation}\label{1.7in3.11}
f^*_{q\Omega'}f_{q\Omega}=\frac{\pi}{k^2}\sum_{\ell ll'}(2l+1)(2l'+1)
\begin{pmatrix} l' & \ell & l \\ 0 & 0 & 0
\end{pmatrix}
^2S^{l\Omega'*}_{l'\ell}(q,\varepsilon)
S^{l\Omega}_{l'\ell}(q,\varepsilon)~,
\end{equation}
The PI cross section, $\sigma^{\mathrm{PI}}({\bf a}_\Omega)$,
obtained after integrating over the energy of the emitted electron and
summing over all exit channels, is,
\begin{equation}
\sigma^{\mathrm {PI}}({\bf a}_\Omega)=\sum_{\Omega'\Omega}{\bf a}^*_{\Omega'}
{\bf a}_{\Omega}\sigma^{\mathrm{PI}}_{\Omega'\Omega}.
\end{equation}

As a first example we examine coherent control obtained using only two
$\Omega(=0,1)$ states [similar results were obtained for the
$\Omega(=0,2)$ pair, as shown in the tables below], for which,
\begin{equation}
\label{1.7in1.13}
\sigma^{\mathrm{PI}}({\bf a}_\Omega)=\left|a_0\right|^2
\sigma^{\mathrm{PI}}_0+\left|a_1\right|^2
\sigma^{\mathrm{PI}}_1+2\Re{\left(a^*_0a_1
\sigma_{01}^{\mathrm{PI}}\right)},
\end{equation}
where $\sigma^{\mathrm{PI}}_\Omega=\sigma^{\mathrm{PI}}_{\Omega\Omega}$.

Similar expressions for AI are obtained
by summing over the exit channels and bound states,
\begin{equation}
\label{1.19.in1}
\sigma^{\mathrm{AI}}=\left|a_0\right|^2\sigma_0^{\mathrm{AI}}+
\left|a_1\right|^2\sigma_1^{\mathrm{AI}}
+2\Re{\left(a^*_0a_1\sigma_{01}^{\mathrm{AI}}\right)}.
\end{equation}
Note the crucial interference term, dependent on the magnitude and
phase of the $a_i$, which allows control over the cross sections by
varying these coefficients.

{\it Computational Results:} Although results are reported for collisions
at temperatures up to 1 K, our main focus is on cold collisions at a temperature
of 1\,mK and on ultracold collisions at 1\,$\mu$K. 
At the low temperatures considered, the PI or AI cross sections are very large
since the two atoms are in the vicinity of one another for an extended
period of time. Further, in the ultracold case only the $s$ partial wave contributes,
with three angular momentum states contributing in the cold case.

For these energies the
relative velocities between the collisional pair are $\approx$ 1\,m/s and
$\approx$ 0.006\,m/s respectively. 
These relative velocities are
experimentally attainable using laser cooling and manipulation techniques.
For example, the atoms can be cooled and trapped in a 3D optical lattice,
then adiabatically accelerated along a single axis \cite{Treutlein01}.
This setup can reach velocities of up to a few meters per second, and a
kinetic energy spread of 150-200\,nK around the central beam velocity. The
present collisional scenario would require two 3D lattice setups in order
to control both scattering particles. The internal state superposition
can be prepared\cite{bergmann06} after cooling while the atoms are trapped in the lattice, for this two techniques are possible: stimulated Raman adiabatic passage (STIRAP) \cite{Bergmann98} or coherent population trapping (CPT) \cite{Kitching01}.

\begin{figure}[p]
\centering
\includegraphics[height=7in]{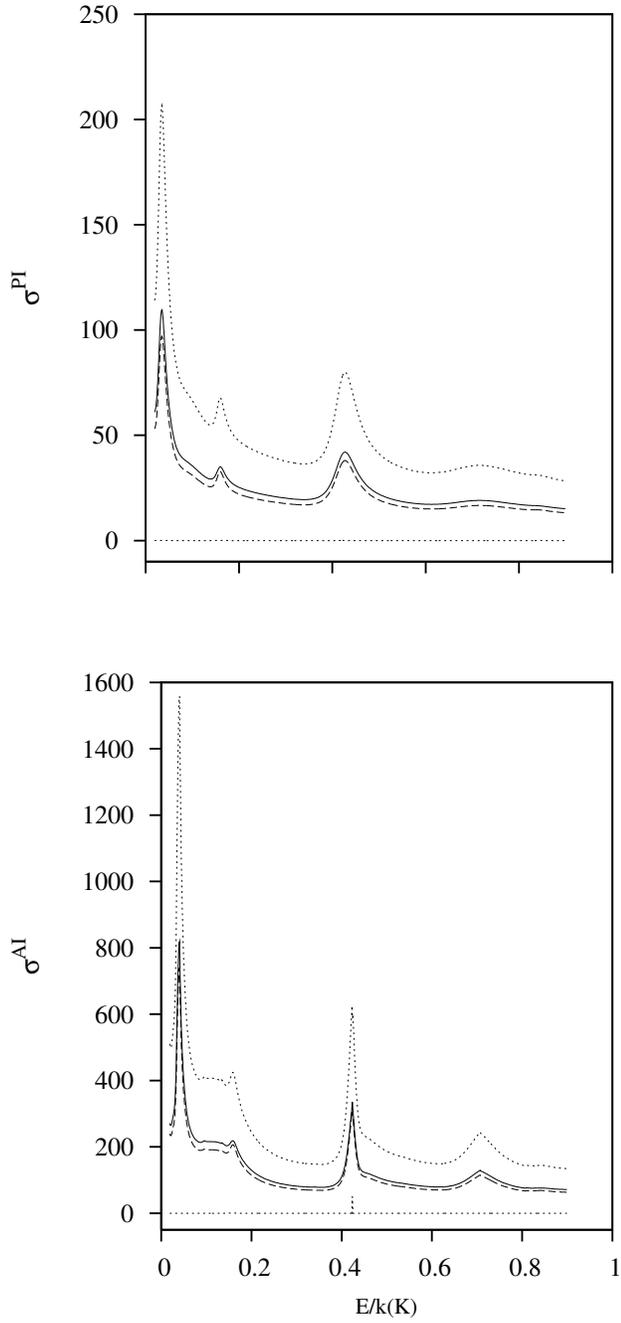}
\caption{\label{fig1}   PI   and   AI   cross   sections   for
$\Omega=0,1$    linear    combination.    $\Omega=0$   -solid;
$\Omega=1$   -dashed;  maximum  and  minimum  for  the  linear
combination of $\Omega=0,1$ -dotted.}
\end{figure}

Consider first control results across a broad spectrum of energies up to
temperatures of 1 K, 
Figure \ref{fig1}  shows the cross sections for $\Omega=0$
and $\Omega=1$ scattering as a function of energy. 
Also
shown are the maximum and minimum controlled 
cross sections (optimized over the $a_i$
coefficients) at each energy for the $\Omega=0$ plus $\Omega=1$ linear
combination. Several resonances\cite{Westphal96} are evident, since the collision energy is
very close to the dissociation threshold for the Ne$^*$-Ar quasimolecule.
Noteworthy is that control is
extensive, with enhancement and suppression of both cross sections being
possible at both resonant and nonresonant energies.

\begin{figure}[p]
\centering
\includegraphics[height=4.5in]{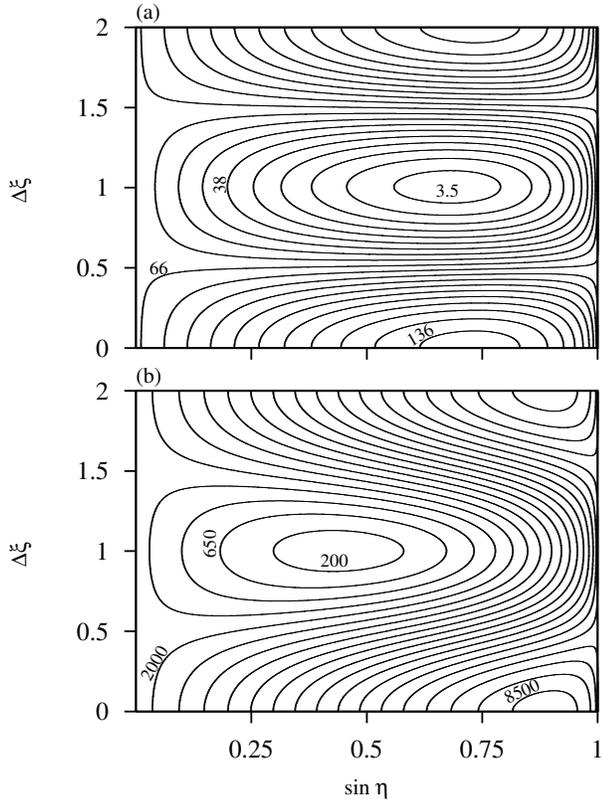}
\caption{\label{fig3}  Coherent  Control  contours  for (a) PI for
$\Omega=0,1$  in  cold  collisions at $T=1\,\mathrm{mK}$; (b)   AI   for
$\Omega=0,2$  in  ultra  cold  collisions  at $T=1\,\mu\mathrm{K}$. 
The parameters $\eta$ and $\Delta\xi$ are defined via: 
$a_\Omega=\sin\eta\,{\mathrm e}^{{\mathrm i}\xi_\Omega}$ and 
$a_{\Omega'}=\cos\eta\,{\mathrm e}^{{\mathrm i}\xi_{\Omega'}}$, 
with $\Omega,\Omega'=0,1,2$.}
\end{figure}

Table \ref{tab2} presents numerical results for the 
cold collision (1\,mK) case. We see that it is possible to
actively change the AI and PI cross sections by as much as four orders of
magnitude for the $\Omega=0,1$ linear combination and three orders of
magnitude for the $\Omega=0,2$ linear combination. For both linear
combinations the position of the minima and maxima for
$\sigma^\mathrm{PI}$ and $\sigma^\mathrm{AI}$ occur at close points in the
parameter space (not shown here). A similar observation has been noted in the thermal case
(above 1\,K), indicating that both PI and AI cross sections can be controlled
simultaneously \cite{Arango06}. Sample results for the control of
$\sigma^{\mathrm{PI}}$ as a function of $a_i$ for the cold collision case
are shown in Fig. \ref{fig3}a.

Ultracold collisions, where only $s$ waves contribute to the process,
show even more dramatic behavior. As seen in Table \ref{tab3}, 
active changes of
up to four orders of magnitude, using the $\Omega=0,1$ superposition
states, and up to three orders of magnitude, using the $\Omega=0,2$
superposition states, are possible. The AI process can also be almost as
well controlled. The resulting $\sigma^{\mathrm{AI}}$ cross sections are
shown in Fig. \ref{fig3}b for ultracold collisions as function of the
$a_i$. Note that in all cases, the maxima and minima in the control plots
(Figs. \ref{fig1} and \ref{fig3}) are well separated, making the
experiment less sensitive to the control parameters.

In summary, we have shown the possibility of a huge range of control of
the PI and AI cross sections in Ne$^*$+Ar cold and ultracold
collisions. Control is achieved by initiating the collision in a
judiciously chosen superposition of Ne$^*$ quantum states. Such states can
be readily made using new STIRAP techniques\cite{bergmann06}. Our results
show a wide range of controllability for both PI and AI. For PI the
minimum of the cross section is found to be orders of magnitude smaller
than the incoherent mixture of $\Omega=0,1$ or $\Omega=0,2$. In the AI
reaction the effects are even more dramatic, showing a minimum four orders
of magnitude smaller than that of the incoherent mixture $\Omega=0,1$ and
three orders of magnitude smaller than the $\Omega=0,2$ mixture.

{\it Acknowledgments:} We thank Professor Peter Siska for making his
computer programs available to us, Professor Klaas Bergmann for extensive
discussions, and Dr. Michael Spanner for his design of optical lattice
implementations of this scenario.

\bibliography{cold}

\begin{thebibliography}{24}
\expandafter\ifx\csname natexlab\endcsname\relax\def\natexlab#1{#1}\fi
\expandafter\ifx\csname bibnamefont\endcsname\relax
  \def\bibnamefont#1{#1}\fi
\expandafter\ifx\csname bibfnamefont\endcsname\relax
  \def\bibfnamefont#1{#1}\fi
\expandafter\ifx\csname citenamefont\endcsname\relax
  \def\citenamefont#1{#1}\fi
\expandafter\ifx\csname url\endcsname\relax
  \def\url#1{\texttt{#1}}\fi
\expandafter\ifx\csname urlprefix\endcsname\relax\def\urlprefix{URL }\fi
\providecommand{\bibinfo}[2]{#2}
\providecommand{\eprint}[2][]{\url{#2}}

\bibitem[{\citenamefont{Shapiro and Brumer}(2003)}]{Shapiro03}
\bibinfo{author}{\bibfnamefont{M.}~\bibnamefont{Shapiro}} \bibnamefont{and}
  \bibinfo{author}{\bibfnamefont{P.}~\bibnamefont{Brumer}},
  \emph{\bibinfo{title}{{Principles of the Quantum Control of Molecular
  Processes}}} (\bibinfo{publisher}{John Willey \& Sons},
  \bibinfo{address}{Hoboken, NJ}, \bibinfo{year}{2003}).

\bibitem[{\citenamefont{Shapiro and Brumer}(1996)}]{Shapiro96}
\bibinfo{author}{\bibfnamefont{M.}~\bibnamefont{Shapiro}} \bibnamefont{and}
  \bibinfo{author}{\bibfnamefont{P.}~\bibnamefont{Brumer}},
  \bibinfo{journal}{Phys. Rev. Lett.} \textbf{\bibinfo{volume}{77}},
  \bibinfo{pages}{2574} (\bibinfo{year}{1996}).

\bibitem[{\citenamefont{Brumer et~al.}(1999)\citenamefont{Brumer, Abrashkevich,
  and Shapiro}}]{Brumer99}
\bibinfo{author}{\bibfnamefont{P.}~\bibnamefont{Brumer}},
  \bibinfo{author}{\bibfnamefont{A.}~\bibnamefont{Abrashkevich}},
  \bibnamefont{and} \bibinfo{author}{\bibfnamefont{M.}~\bibnamefont{Shapiro}},
  \bibinfo{journal}{Faraday Discuss.} \textbf{\bibinfo{volume}{113}},
  \bibinfo{pages}{291} (\bibinfo{year}{1999}).

\bibitem[{\citenamefont{Rice and Zhao}(2000)}]{Rice}
\bibinfo{author}{\bibfnamefont{S.}~\bibnamefont{Rice}} \bibnamefont{and}
  \bibinfo{author}{\bibfnamefont{M.}~\bibnamefont{Zhao}},
  \emph{\bibinfo{title}{{Optical Control of Molecular Dynamics}}}
  (\bibinfo{publisher}{John Willey \& Sons}, \bibinfo{address}{Hoboken, NJ},
  \bibinfo{year}{2000}).

\bibitem[{\citenamefont{Siska}(1993)}]{Siska93}
\bibinfo{author}{\bibfnamefont{P.~E.} \bibnamefont{Siska}},
  \bibinfo{journal}{Rev. Mod. Phys.} \textbf{\bibinfo{volume}{65}},
  \bibinfo{pages}{337} (\bibinfo{year}{1993}).

\bibitem[{\citenamefont{Arango et~al.}(2006)\citenamefont{Arango, Shapiro, and
  Brumer}}]{Arango06}
\bibinfo{author}{\bibfnamefont{C.~A.} \bibnamefont{Arango}},
  \bibinfo{author}{\bibfnamefont{M.}~\bibnamefont{Shapiro}}, \bibnamefont{and}
  \bibinfo{author}{\bibfnamefont{P.}~\bibnamefont{Brumer}},
  \bibinfo{journal}{J. Chem. Phys.} \textbf{\bibinfo{volume}{in press}},
  \bibinfo{pages}{xxx} (\bibinfo{year}{2006}).

\bibitem[{\citenamefont{Mori et~al.}(1964)\citenamefont{Mori, Watanabe, and
  Fujita}}]{Mori64}
\bibinfo{author}{\bibfnamefont{M.}~\bibnamefont{Mori}},
  \bibinfo{author}{\bibfnamefont{T.}~\bibnamefont{Watanabe}}, \bibnamefont{and}
  \bibinfo{author}{\bibfnamefont{H.}~\bibnamefont{Fujita}},
  \bibinfo{journal}{J. Phys. Soc. Jap.} \textbf{\bibinfo{volume}{19}},
  \bibinfo{pages}{380} (\bibinfo{year}{1964}).

\bibitem[{\citenamefont{Hickman and Morgner}(1976)}]{Hickman76}
\bibinfo{author}{\bibfnamefont{A.~P.} \bibnamefont{Hickman}} \bibnamefont{and}
  \bibinfo{author}{\bibfnamefont{H.}~\bibnamefont{Morgner}},
  \bibinfo{journal}{J. Phys. B} \textbf{\bibinfo{volume}{9}},
  \bibinfo{pages}{1765} (\bibinfo{year}{1976}).

\bibitem[{\citenamefont{Morgner}(1978)}]{Morgner78}
\bibinfo{author}{\bibfnamefont{H.}~\bibnamefont{Morgner}}, \bibinfo{journal}{J.
  Phys. B.} \textbf{\bibinfo{volume}{11}}, \bibinfo{pages}{269}
  (\bibinfo{year}{1978}).

\bibitem[{\citenamefont{Bieniek}(1978)}]{Bieniek78}
\bibinfo{author}{\bibfnamefont{R.~J.} \bibnamefont{Bieniek}},
  \bibinfo{journal}{Phys. Rev. A} \textbf{\bibinfo{volume}{18}},
  \bibinfo{pages}{392} (\bibinfo{year}{1978}).

\bibitem[{\citenamefont{O'Malley}(1966)}]{O'Malley66}
\bibinfo{author}{\bibfnamefont{T.~F.} \bibnamefont{O'Malley}},
  \bibinfo{journal}{Phys. Rev.} \textbf{\bibinfo{volume}{150}},
  \bibinfo{pages}{14} (\bibinfo{year}{1966}).

\bibitem[{\citenamefont{O'Malley}(1967)}]{O'Malley67}
\bibinfo{author}{\bibfnamefont{T.~F.} \bibnamefont{O'Malley}},
  \bibinfo{journal}{Phys. Rev.} \textbf{\bibinfo{volume}{156}},
  \bibinfo{pages}{230} (\bibinfo{year}{1967}).

\bibitem[{\citenamefont{Miller et~al.}(1972)\citenamefont{Miller, Slocomb, and
  SchaeferIII}}]{Miller72}
\bibinfo{author}{\bibfnamefont{W.~H.} \bibnamefont{Miller}},
  \bibinfo{author}{\bibfnamefont{C.~A.} \bibnamefont{Slocomb}},
  \bibnamefont{and} \bibinfo{author}{\bibfnamefont{H.~F.}
  \bibnamefont{SchaeferIII}}, \bibinfo{journal}{J. Chem. Phys.}
  \textbf{\bibinfo{volume}{56}}, \bibinfo{pages}{1347} (\bibinfo{year}{1972}).

\bibitem[{\citenamefont{Gregor and Siska}(1981)}]{Gregor81}
\bibinfo{author}{\bibfnamefont{R.~W.} \bibnamefont{Gregor}} \bibnamefont{and}
  \bibinfo{author}{\bibfnamefont{P.~E.} \bibnamefont{Siska}},
  \bibinfo{journal}{J. Chem. Phys.} \textbf{\bibinfo{volume}{74}},
  \bibinfo{pages}{1078} (\bibinfo{year}{1981}).

\bibitem[{\citenamefont{Verheijen and Beijerinck}(1986)}]{Verheijen86}
\bibinfo{author}{\bibfnamefont{M.~J.} \bibnamefont{Verheijen}}
  \bibnamefont{and} \bibinfo{author}{\bibfnamefont{H.~C.~W.}
  \bibnamefont{Beijerinck}}, \bibinfo{journal}{Chem. Phys.}
  \textbf{\bibinfo{volume}{102}}, \bibinfo{pages}{255} (\bibinfo{year}{1986}).

\bibitem[{\citenamefont{Kerstel et~al.}(1988)\citenamefont{Kerstel,
  Jan$\ss$ens, van Leeuwen, and Beijerinck}}]{Kerstel88}
\bibinfo{author}{\bibfnamefont{E.~R.~T.} \bibnamefont{Kerstel}},
  \bibinfo{author}{\bibfnamefont{M.~F.~M.} \bibnamefont{Jan$\ss$ens}},
  \bibinfo{author}{\bibfnamefont{K.~A.~H.} \bibnamefont{van Leeuwen}},
  \bibnamefont{and} \bibinfo{author}{\bibfnamefont{H.~C.~W.}
  \bibnamefont{Beijerinck}}, \bibinfo{journal}{Chem. Phys.}
  \textbf{\bibinfo{volume}{119}}, \bibinfo{pages}{325} (\bibinfo{year}{1988}).

\bibitem[{\citenamefont{de~Beek et~al.}(1997)\citenamefont{de~Beek,
  Drie$\ss$en, Kokkelmans, Boom, Biejerinck, and Varhaar}}]{OpdeBeek97}
\bibinfo{author}{\bibfnamefont{S.~S.~O.} \bibnamefont{de~Beek}},
  \bibinfo{author}{\bibfnamefont{J.~P.~J.} \bibnamefont{Drie$\ss$en}},
  \bibinfo{author}{\bibfnamefont{S.~J. J. M.~F.} \bibnamefont{Kokkelmans}},
  \bibinfo{author}{\bibfnamefont{W.}~\bibnamefont{Boom}},
  \bibinfo{author}{\bibfnamefont{H.~C.~W.} \bibnamefont{Biejerinck}},
  \bibnamefont{and} \bibinfo{author}{\bibfnamefont{B.~J.}
  \bibnamefont{Varhaar}}, \bibinfo{journal}{Phys. Rev. A}
  \textbf{\bibinfo{volume}{56}}, \bibinfo{pages}{2792} (\bibinfo{year}{1997}).

\bibitem[{\citenamefont{Morgner}(1982)}]{Morgner82}
\bibinfo{author}{\bibfnamefont{H.}~\bibnamefont{Morgner}},
  \bibinfo{journal}{Comments At. Mol. Phys.} \textbf{\bibinfo{volume}{11}},
  \bibinfo{pages}{271} (\bibinfo{year}{1982}).

\bibitem[{\citenamefont{Morgner}(1985)}]{Morgner85}
\bibinfo{author}{\bibfnamefont{H.}~\bibnamefont{Morgner}}, \bibinfo{journal}{J.
  Phys. B: At. Mol. Phys.} \textbf{\bibinfo{volume}{18}}, \bibinfo{pages}{251}
  (\bibinfo{year}{1985}).

\bibitem[{\citenamefont{Treutlein et~al.}(2001)\citenamefont{Treutlein, Chung,
  and Chu}}]{Treutlein01}
\bibinfo{author}{\bibfnamefont{P.}~\bibnamefont{Treutlein}},
  \bibinfo{author}{\bibfnamefont{K.~Y.} \bibnamefont{Chung}}, \bibnamefont{and}
  \bibinfo{author}{\bibfnamefont{S.}~\bibnamefont{Chu}},
  \bibinfo{journal}{Phys. Rev. A} \textbf{\bibinfo{volume}{63}},
  \bibinfo{pages}{051401} (\bibinfo{year}{2001}).

\bibitem[{\citenamefont{Heinz et~al.}(2006)\citenamefont{Heinz, Vewinger,
  Schneider, Yatsenko, and Bergmann}}]{bergmann06}
\bibinfo{author}{\bibfnamefont{M.}~\bibnamefont{Heinz}},
  \bibinfo{author}{\bibfnamefont{F.}~\bibnamefont{Vewinger}},
  \bibinfo{author}{\bibfnamefont{U.}~\bibnamefont{Schneider}},
  \bibinfo{author}{\bibfnamefont{L.}~\bibnamefont{Yatsenko}}, \bibnamefont{and}
  \bibinfo{author}{\bibfnamefont{K.}~\bibnamefont{Bergmann}},
  \bibinfo{journal}{Opt. Comm.} \textbf{\bibinfo{volume}{264}},
  \bibinfo{pages}{248} (\bibinfo{year}{2006}).

\bibitem[{\citenamefont{Bergmann et~al.}(1998)\citenamefont{Bergmann, H., and
  B.W.}}]{Bergmann98}
\bibinfo{author}{\bibfnamefont{K.}~\bibnamefont{Bergmann}},
  \bibinfo{author}{\bibfnamefont{T.}~\bibnamefont{H.}}, \bibnamefont{and}
  \bibinfo{author}{\bibfnamefont{S.}~\bibnamefont{B.W.}},
  \bibinfo{journal}{Rev. Mod. Phys.} \textbf{\bibinfo{volume}{70}},
  \bibinfo{pages}{1003} (\bibinfo{year}{1998}).

\bibitem[{\citenamefont{J.~Kitching et~al.}(2001)\citenamefont{J.~Kitching,
  Hollberg, Wynanads, and Knappe}}]{Kitching01}
\bibinfo{author}{\bibfnamefont{H.~R.} \bibnamefont{J.~Kitching}},
  \bibinfo{author}{\bibfnamefont{L.}~\bibnamefont{Hollberg}},
  \bibinfo{author}{\bibfnamefont{R.}~\bibnamefont{Wynanads}}, \bibnamefont{and}
  \bibinfo{author}{\bibfnamefont{S.}~\bibnamefont{Knappe}},
  \bibinfo{journal}{J. Opt. Soc. Am. B} \textbf{\bibinfo{volume}{18}},
  \bibinfo{pages}{1545} (\bibinfo{year}{2001}).

\bibitem[{\citenamefont{Westphal et~al.}(1996)\citenamefont{Westphal, Horn,
  Koch, Schmand, and Andr{\"a}}}]{Westphal96}
\bibinfo{author}{\bibfnamefont{P.}~\bibnamefont{Westphal}},
  \bibinfo{author}{\bibfnamefont{A.}~\bibnamefont{Horn}},
  \bibinfo{author}{\bibfnamefont{S.}~\bibnamefont{Koch}},
  \bibinfo{author}{\bibfnamefont{J.}~\bibnamefont{Schmand}}, \bibnamefont{and}
  \bibinfo{author}{\bibfnamefont{H.~J.} \bibnamefont{Andr{\"a}}},
  \bibinfo{journal}{Phys. Rev. A} \textbf{\bibinfo{volume}{54}},
  \bibinfo{pages}{4577} (\bibinfo{year}{1996}).

\end{thebibliography}

\begin{center}
\begin{table}[p]
\caption{\label{tab2} Cross section for cold collision at T = 1\,mK. Rows
labelled ``$\Omega=0,1$" and ``$\Omega=0,2$" show the minimum and maximum
of the $\sigma^\mathrm{PI}$ and $\sigma^\mathrm{AI}$, obtained by varying
the $a_i$ for the indicated superposition.}
\begin{center}
\begin{tabular}{|c|c|c|}
\hline
                &                                               &     \\
~~~$\Omega$~~~  &  ~~~$\sigma^\mathrm{PI}$($\mathrm{\AA}^2$)~~~ &    ~~~$\sigma^\mathrm{AI}$($\mathrm{\AA}^2$)~~~ \\
                &                                               &     \\
\hline
0   &  74.68                             &  346.91         \\
1   &  64.90                             &  306.25         \\
2   &  13.75                             &  87.01          \\
0,1 &  1.27$\times 10^{-2}$ $-$ 139.57   &  $3\times 10^{-2}$  $-$ 653.13\\
0,2 &   0.63 $-$ 87.80 &  0.60 $-$ 433.32   \\
\hline
\end{tabular}
\end{center}
\end{table}
\end{center}

\newpage

\begin{center}
\begin{table}[p]
\caption{\label{tab3} As in Table I, but for ultra cold collision at T =
1\,$\mu$K.}
\begin{center}
\begin{tabular}{|c|c|c|}
\hline
                &                                             &     \\
~~~$\Omega$~~~  &  ~~~$\sigma^\mathrm{PI}$($\mathrm{\AA}^2$)~~~ &    ~~~$\sigma^\mathrm{AI}$($\mathrm{\AA}^2$)~~~ \\
                &                                             &     \\
\hline
0   &  1357.32                           & 7056.92         \\
1   &  1174.19                           & 6199.40         \\
2   &  244.27                           &  1728.65          \\
0,1 &  0.128 $-$ 2531.38                &  3.03 $-$ 13256.30 \\
0,2 &  9.88  $-$ 1591.71 & 2.28  $-$  8783.30  \\
\hline
\end{tabular}
\end{center}
\end{table}
\end{center}

\newpage

\end{document}